\documentclass[twocolumn,aps,pra,showpacs,longbibliography,superscriptaddress]{revtex4-1}

\usepackage{amssymb,latexsym}
\usepackage{mathrsfs}
\usepackage{lipsum}
\usepackage{graphicx}
\usepackage[caption=false]{subfig}
\usepackage{mathtools}
\usepackage{epstopdf}
\usepackage{xcolor}
\usepackage[colorlinks, linkcolor=red, anchorcolor=green, citecolor=green, urlcolor=blue]{hyperref}
\DeclareGraphicsExtensions{.pdf,.jpg,.png,.eps}
\usepackage[toc,page,title,titletoc,header]{appendix}
\usepackage{etoolbox}
\usepackage{breqn}

\makeatletter
\let\cat@comma@active\@empty
\makeatother

\begin{document}
	
\title{Performance advantage of discriminating one-versus-two incoherent sources based on quantum hypothesis testing}

\author{Jian-Dong Zhang}
\email[]{zhangjiandong1993@gmail.com}
\affiliation{School of Mathematics and Physics, Jiangsu University of Technology, Changzhou 213001, China}
\author{Mei-Ming Zhang}
\affiliation{School of Mathematics and Physics, Jiangsu University of Technology, Changzhou 213001, China}
\author{Chuang Li}
\affiliation{Research Center for Novel Computing Sensing and Intelligent Processing, Zhejiang Lab, Hangzhou 311121, China}
\author{Shuai Wang}
\affiliation{School of Mathematics and Physics, Jiangsu University of Technology, Changzhou 213001, China}	
\date{\today}
	
\begin{abstract}
Detecting the presence of multiple incoherent sources is a fundamental and challenging task for quantum imaging, especially within sub-Rayleigh region.	
In this paper, the discrimination of one-versus-two point-like incoherent sources in symmetric and asymmetric scenarios is studied. 
We calculate the quantum lower bounds on error probabilities of making a decision after one-shot and multi-shot tests.
The results are compared with the error probability of prior-based direct guess, and the minimal number of tests required to make a decision outperforming direct guess is discussed.
We also show the asymptotic quantum lower bound for a large number of tests.
For practical purposes, we propose a specific strategy along with decision rule of which can work without any prior knowledge.
With respect to each of two scenarios, the error probability can approach the quantum lower bound in one-shot test as well as multi-shot test. 
In addition, the potential challenges and solutions in a realistic scenario are analyzed.
Our results may contribute to real-world quantum imaging such as microscopy and astronomy.
\end{abstract}

\maketitle

\section{Introduction}
Quantum imaging is an art and science that exploits diverse quantum techniques to break the performance bound given by conventional direct imaging.
In general, the resolution of an optical system is characterized by the ability to discriminate two incoherent point-like sources.
Related to this, one of the most important tasks in imaging is to determine whether an additional incoherent source exists on the image plane.
In this regard, it is indisputable that Rayleigh’s criterion has been the most influential measure.
In terms of the criterion, only when two incoherent sources are separated at least by a
diffraction-limited size on the image plane can we achieve perfect discrimination.
In fact, the Rayleigh's criterion does not provide a quantitative measure of distinguishability and is mainly aimed at direct imaging or vision perception.
Interests in other methods which can achieve better resolution are triggered by important significance in various practical applications.

With a comprehensive analysis on the quantum Fisher information, Tsang \emph{et al.} pointed out that the Rayleigh's curse can be broken through the use of quantum technology \cite{PhysRevX.6.031033}.
To be specific, two incoherent sources can be resolved with constant precision limit, irrespective of their spatial separation.
Since then, a lot of attentions have focused on the estimation of the spatial separation between the two sources. 
In order to achieve the optimal estimation precision given by the quantum Fisher information, some feasible methods were proposed, such as spatial-mode
demultiplexing (SPADE) \cite{PhysRevA.97.023830,Tsang_2017}, super-localization via image-inversion
interferometry (SLIVER) \cite{Nair:16} and super-resolved position localisation by
inversion of coherence along an edge (SPLICE) \cite{PhysRevLett.118.070801,Bonsma-Fisher_2019}.
The relevant proof-of-principle experiments were also reported \cite{Paur:16,Tang:16,Larson2019,Santamaria:24}.
Meanwhile, estimation of two incoherent sources with unequal brightnesses was discussed \cite{PhysRevA.96.062107,PhysRevA.98.012103,PhysRevA.103.052604,Prasad_2020}.
The estimation precision in one-dimensional to three-dimensional scenarios were also analyzed \cite{Zhang:24,PhysRevA.95.063847,Zhou:19,PhysRevLett.121.180504,PhysRevA.99.022116,Wang:21}.
More recently, the methods to calculate the precision limit of more than two point sources were proposed \cite{Bisketzi_2019,PhysRevLett.124.080503}.
In addition, some studies focused on practical scenarios including partially coherent sources \cite{Larson:18,Tsang:2019,Larson:19,Wadood:19} and noisy environment or detector \cite{PhysRevLett.126.120502,doi:10.1142/S0219749919410156,PhysRevA.101.022323}.

Aside from parameter estimation, the task of discriminating one-versus-two incoherent sources can be also investigated by using the tool of hypothesis testing.
The goal is to minimize the error probability.
So far, the relevant studies have focused on the quantum lower bound on the error probability and the methods approaching this bound \cite{PhysRevA.103.022406,zanforlin2022optical,PhysRevLett.127.130502,lu2018quantum,Wadood24}.
These results are helpful complement to the theory of incoherent imaging, and the combination of parameter estimation and hypothesis testing will promote the implementation of versatile quantum imaging. 
However, there are still some issues that remain to be addressed. 
On the one hand, these studies only analyzed two hypotheses with equal prior probabilities, which are maybe not applicable in some cases.
On the other hand, the important performance metric, quantum-optimal error probability with different parameters in one-shot test, is yet unclear.

In this paper, we revisit the above task and consider two hypotheses with unequal prior probabilities.
The quantum-optimal error probabilities in symmetric and asymmetric scenarios are calculated in terms of the theory of quantum hypothesis testing.
With respect to one-shot test, we analyze the performance advantage of the quantum-optimal error probability over the error probability of direct guess based on prior information.
For multi-shot test, we study the performance advantage of few-shot test and the asymptotic quantum-optimal error probability of a large number of tests.
In particular, we show a specific strategy and decision rule, which can nearly saturate the quantum-optimal error probability either in one-shot test or multi-shot test.
These results are guidance for practical applications.

The remainder of this paper is organized as follows.
Section \ref{s2} introduces the fundamental model and quantum lower bounds in symmetric and asymmetric scenarios.
In Secs. \ref{s3} and \ref{s4}, we calculate and analyze the quantum-optimal error probabilities in one-shot test and multi-shot test, respectively.
In Sec. \ref{s5}, we propose a specific strategy along with decision rule and compare its error probability with the quantum-optimal error probability.
The potential challenges and solutions in a realistic scenario are discussed in Sec. \ref{s6}.
Finally, we summarize our main results in Sec. \ref{s7}.

\section{Fundamental theory}
\label{s2}

\begin{figure}[htbp]
	\centering	\includegraphics[width=0.48\textwidth]{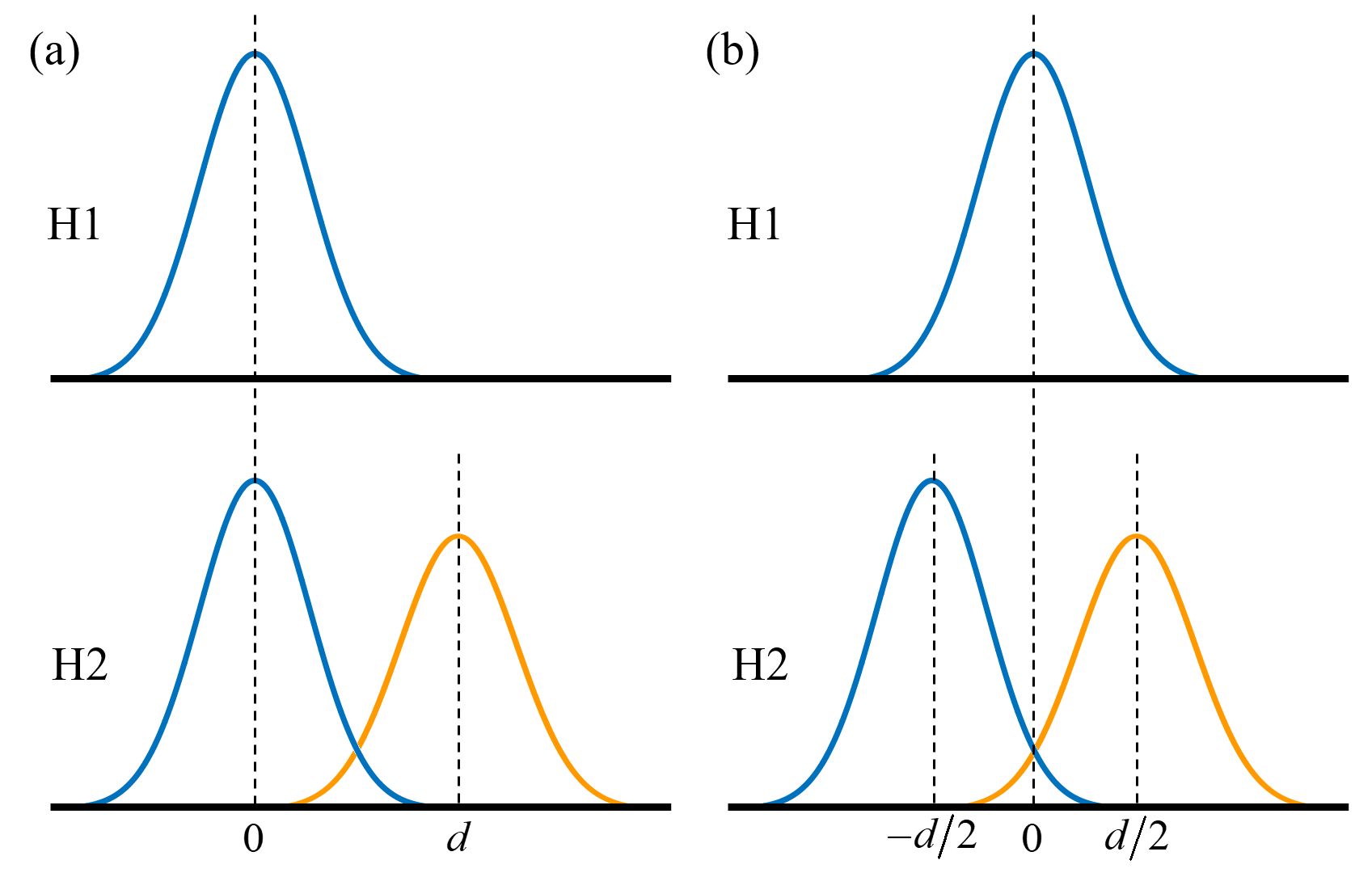}
	\caption{Schematic of (a) asymmetric scenario and (b) symmetric scenario of incoherent sources on the image plane. The spatial separation between the two incoherent sources is $d$. The point spread function of optical system follows Gaussian distribution.}
	\label{system}
\end{figure}

We first lay out the basic model of incoherent source discrimination used in this paper, as shown in Fig. \ref{system}.
The task is to determine whether an additional incoherent 
source exists. 
The result can be 
classified into $H_1$ hypothesis, a single source, and $H_2$ hypothesis, two incoherent sources. 
Figures \ref{system}(a) and \ref{system}(b) show asymmetric and symmetric scenarios, respectively.
In an asymmetric scenario, the first source is fixed at 0 without moving under $H_1$ and $H_2$ hypotheses, while the second source appears at $d$ under $H_2$ hypothesis.
In a symmetric scenario, the first source sits at $0$ under $H_1$ hypothesis, while under $H_2$ hypothesis two sources are symmetrically disposed about 0.
Without loss of generality, here we assume that two sources are of unequal brightnesses.

In terms of quantum theory, two hypotheses--one source versus two incoherent sources--can be denoted by two quantum states $\Lambda _1$ and $\Lambda _2$, respectively.
Of the most applications in microscopy and astronomy, total source brightness is faint, i.e., $\epsilon \ll 1$.
For this reason, high-photon states can be ignored and the quantum states may be regarded as the superposition of zero-photon and one-photon states,
\begin{align}
\Lambda _ {i} = (1-\epsilon)\eta + \epsilon \rho_i
\label{e1}
\end{align}
with $i = 1,2$, where $\eta$ and $\rho$ are the density matrices of zero-photon and one-photon states.
Under two hypotheses, only one-photon states are different as zero-photon states are uninformative.

Generally, we have no information on the total source brightness and the ratio of zero-photon states to one-photon states is unknown.
In this situation, a practicable method is to merely consider one-photon states.
To be specific, only when the detector is triggered by a single photon can we accept it as an effective detection event.
Hence, two quantum states $\Lambda _1$ and $\Lambda _2$ reduce to $\rho _1$ and $\rho _2$.
Throughout this paper, a single effective detection event is defined as one-shot test, and $M$-shot (multi-shot) test refers to $M$ sequential effective detection events.
Related to this, one can make a decision between the two hypotheses after one-shot or multi-shot test.

We assume that the prior probabilities of two hypotheses are $P_1$ and $P_2$.
The quantum lower bound on the error probability, the quantum-optimal error probability, of making a decision after one-shot test is found to be \cite{1055052,HELSTROM1967254} 
\begin{align}
E_{\min} = \min \left[ {{P_{\rm{err}}}} \right] = \frac{1}{2}\left( {1 - \left\| {{P_2}{\rho _2} - {P_1}{\rho _1}}\right\|} \right), 
\label{e2}
\end{align}
where $\left\| \cdot  \right\|$ denotes the trace norm (sum of singular values) of the matrix.
This lower bound is known as the quantum Helstrom bound. 
Since ${{P_2}{\rho _2} - {P_1}{\rho _1}}$ is Hermitian, the trace norm is equal to the sum of absolute values of the eigenvalues \cite{PhysRevA.71.062340}.
For a decision after $M$-shot test, the quantum Helstrom bound is given by
\begin{align}
{E_{\min }} = \min \left[ {{P_{{\rm{err}}}}} \right] = \frac{1}{2}\left( {1 - \left\| {{P_2}\rho _2^{\otimes M} - {P_1}\rho _1^{\otimes M}} \right\|} \right).
\label{e3}
\end{align}
This result reduces to Eq. (\ref{e2}) with $M = 1$.

When two prior probabilities are unequal, prior-based direct guess is a convenient and costless strategy.
The error probability of direct guess is given by $E_{\rm g} = \min\left\{ P_1, P_2 \right\}$. 
We have $E_{\min} \le E_{\rm g}$ and $E_{\min} = E_{\rm g}$ is obtained when all eigenvalues of or ${{P_2}\rho _2^{\otimes M} - {P_1}\rho _1^{\otimes M}} $ are positive or negative.
As a result, $A = E_{\rm g}/E_{\min}$ can be regarded as performance advantage.
For a fixed $M$, we may get $E_{\min} = E_{\rm g}$ when the prior probability is above or below a critical value.
At this time, any detection becomes useless, and the optimal strategy is directly guessing “$H_1$” or “$H_2$”. 
That is, any detection cannot provide a lower error probability than direct guess.
Related to this, the region with $E_{\min} = E_{\rm g}$ is regarded as forbidden region.

\section{Quantum bound on one-shot test}
\label{s3}
In this section, we focus on the quantum-optimal error probability of making a decision after one-shot test. 
The relevant results are compared with the error probability of prior-based direct guess.
Both of performance advantage and forbidden region are analyzed.

We first pay our attention to an asymmetric scenario.
At this time, quantum states under two hypotheses can be described as 
\begin{align}
{\rho _1} = \left| {{\psi _0}} \right\rangle \left\langle {{\psi _0}} \right|,
\label{e4}
\end{align}
\begin{align}
{\rho _2} = q\left| {{\psi _0}} \right\rangle \left\langle {{\psi _0}} \right| + \left( {1 - q} \right)\left| {{\psi _d}} \right\rangle \left\langle {{\psi _d}} \right|,
\label{e5}
\end{align}
where $q$ is the weighting related to the relative brightness and the subscript represents the central position of the corresponding source.
The above quantum states only involve two ket vectors, so ${{P_2}{\rho _2} - {P_1}{\rho _1}}$ can be represented by at least a two-dimensional matrix.

It is worth noting that the two ket vectors are normalized but not orthogonal, i.e., $\left\langle {{{\psi _0}}}
{\left | {\vphantom {{{\psi _0}} {{\psi _0}}}}
	\right. \kern-\nulldelimiterspace}
{{{\psi _0}}} \right\rangle = \left\langle {{{\psi _d}}}
{\left | {\vphantom {{{\psi _d}} {{\psi _d}}}}
	\right. \kern-\nulldelimiterspace}
{{{\psi _d}}} \right\rangle = 1$ and $\left\langle {{{\psi _0}}}
{\left | {\vphantom {{{\psi _0}} {{\psi _d}}}}
	\right. \kern-\nulldelimiterspace}
{{{\psi _d}}} \right\rangle \neq 0$. These results can be proven through the use of
\begin{align}
\left\langle {{{\psi _i}}}
{\left | {\vphantom {{{\psi _i}} {{\psi _j}}}}
	\right. \kern-\nulldelimiterspace}
{{{\psi _j}}} \right\rangle  = \int_{ - \infty }^\infty  {\left\langle {{{\psi _i}}}
	{\left | {\vphantom {{{\psi _i}} x}}
		\right. \kern-\nulldelimiterspace}
	{x} \right\rangle \left\langle {x}
	{\left | {\vphantom {x {{\psi _j}}}}
		\right. \kern-\nulldelimiterspace}
	{{{\psi _j}}} \right\rangle dx} 
\label{e6}
\end{align}
with $\left\{ {i,j} \right\} \in \left\{ {0, d } \right\}$, Gaussian-type point-spread function
\begin{align}
\left\langle {x}
{\left | {\vphantom {x {{\psi _0}}}}
	\right. \kern-\nulldelimiterspace}
{{{\psi _0}}} \right\rangle = {\left( {\frac{1}{{2\pi {\sigma ^2}}}} \right)^{{1 \mathord{\left/
				{\vphantom {1 4}} \right.
				\kern-\nulldelimiterspace} 4}}}\exp \left( { - \frac{{{x^2}}}{{4{\sigma ^2}}}} \right),
\label{e7}
\end{align}
\begin{align}
\left\langle {x}
{\left | {\vphantom {x {{\psi _ d }}}}
	\right. \kern-\nulldelimiterspace}
{{{\psi _ d }}} \right\rangle  = {\left( {\frac{1}{{2\pi {\sigma ^2}}}} \right)^{{1 \mathord{\left/
				{\vphantom {1 4}} \right.
				\kern-\nulldelimiterspace} 4}}}\exp \left[ { - \frac{{{{\left( {x -d} \right)}^2}}}{{4{\sigma ^2}}}} \right],
\label{e8}
\end{align}
and integral formula
\begin{align}
\int_{ - \infty }^\infty  {\exp \left( { - a{x^2} + bx + c} \right)dx}  = \sqrt {\frac{\pi }{a}} \exp \left( {\frac{{{b^2}}}{{4a}} + c} \right).
\label{e9}
\end{align}
In order to calculate the trace norm of ${{P_2}{\rho _2} - {P_1}{\rho _1}}$, we reconstruct two orthonormalized ket vectors with the Schmidt's orthogonalization,
\begin{align}
\left| 0 \right\rangle  =& \left| {{\psi _0}} \right\rangle, \\
\left| 1 \right\rangle  =& \frac{1}{{\sqrt {1 - {{\left| \tau  \right|}^2}} }}\left( {\left| {{\psi _d}} \right\rangle  - \tau \left| 0 \right\rangle } \right) 
\label{e11}
\end{align}
with
\begin{align}
\tau  = \left\langle {{{\psi _0}}}
{\left | {\vphantom {{{\psi _0}} {{\psi _d}}}}
	\right. \kern-\nulldelimiterspace}
{{{\psi _d}}} \right\rangle  = \exp \left( { - \frac{{{k^2}}}{{8}}} \right).
\label{e12}
\end{align}
Here $k = d/\sigma$ is the dimensionless parameter denoting the degree of separation between the two sources.

In the representation composed of $\left| 0 \right\rangle$ and $\left| 1 \right\rangle$, two quantum states can be written as
\begin{align}
{\rho _1} = \left[ {\begin{array}{*{20}{c}}
	1&0\\
	0&0
	\end{array}} \right],
\label{e13}
\end{align}
\begin{align}
{\rho _2} = q\left[ {\begin{array}{*{20}{c}}
	1&0\\
	0&0
	\end{array}} \right] + \left( {1 - q} \right)\left[ {\begin{array}{*{20}{c}}
	{{\tau ^2}}&{\tau \sqrt {1 - {\tau ^2}} }\\
	{\tau \sqrt {1 - {\tau ^2}} }&{1 - {\tau ^2}}
	\end{array}} \right].
\label{e14}
\end{align}
Further, one can calculate the quantum-optimal error probability according to Eq. (\ref{e2}).

\begin{figure*}[htbp]
	\centering
	\includegraphics[width=0.3\textwidth]{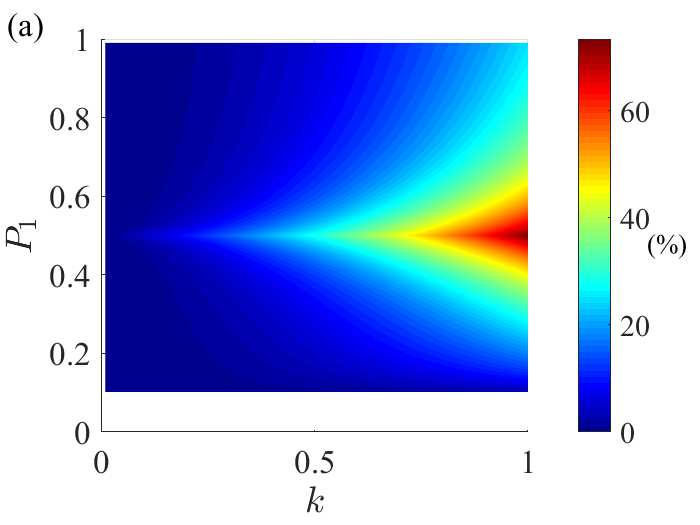}	\includegraphics[width=0.3\textwidth]{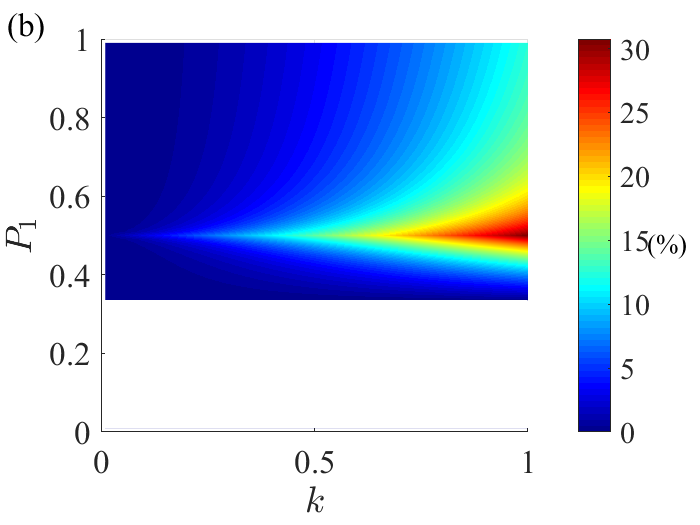}
	\includegraphics[width=0.3\textwidth]{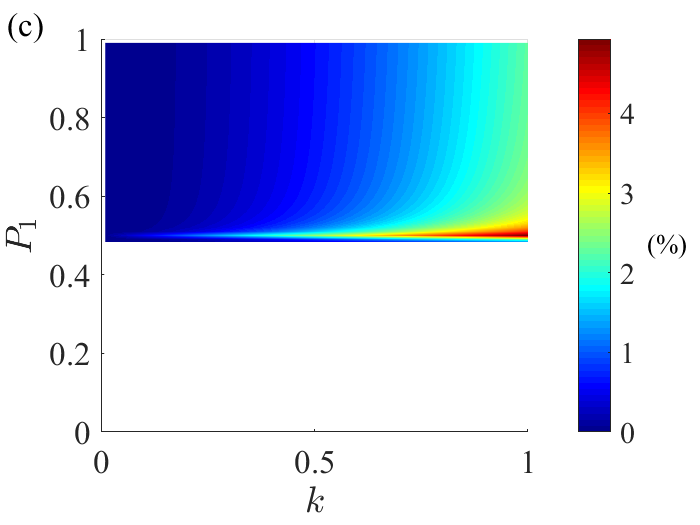}
	\caption{One-shot performance advantage as a function of prior probability and separation in an asymmetric scenario, (a) $q = 0.1$; (b) $q = 0.5$; (c) $q = 0.9$. The blank regions are forbidden regions and the color bars are displayed in percentage form.}
	\label{2Q}
\end{figure*}

\begin{figure*}[htbp]
	\centering
	\includegraphics[width=0.3\textwidth]{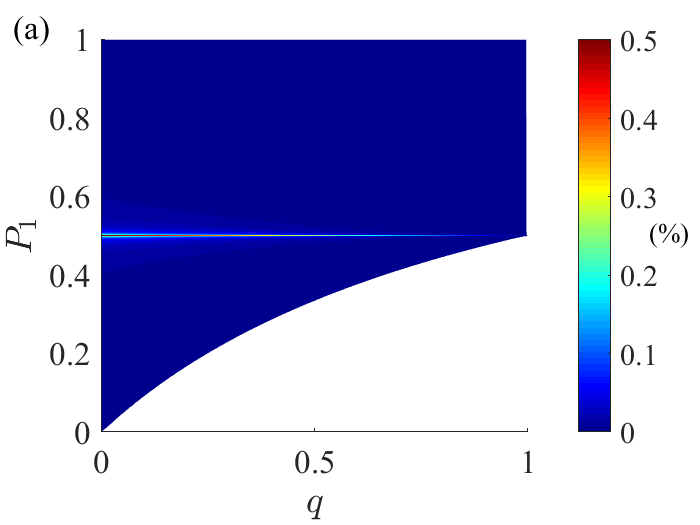}
	\includegraphics[width=0.3\textwidth]{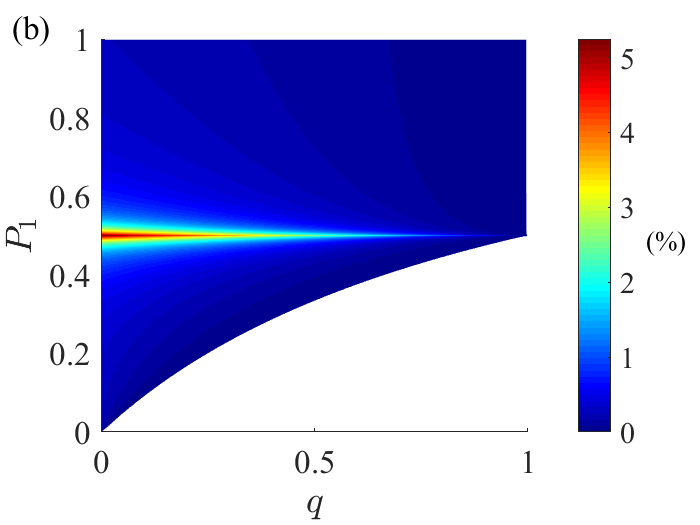}
	\includegraphics[width=0.3\textwidth]{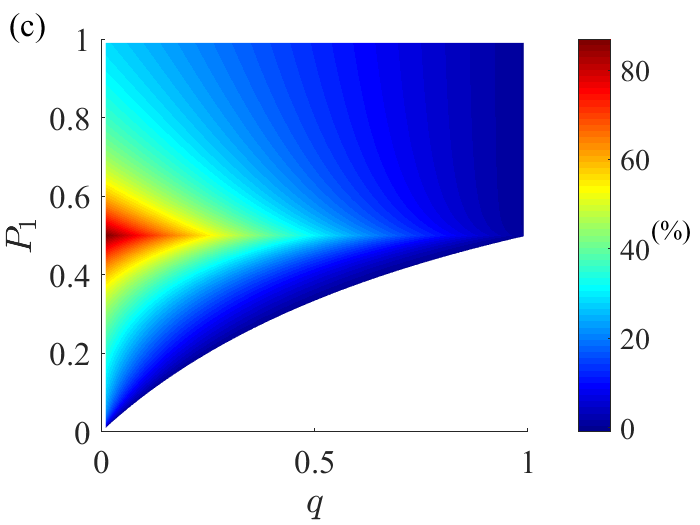}
	\caption{One-shot performance advantage as a function of prior probability and weighting in an asymmetric scenario, (a) $k = 0.01$; (b) $k = 0.1$; (c) $k = 1$. The blank regions are forbidden regions and the color bars are displayed in percentage form.}
	\label{2K}
\end{figure*}

To show the result more intuitively, in Fig. \ref{2Q} we give the dependence of performance advantage on prior probability and separation.
It can be seen that, for a fixed prior probability, the performance advantage increases with the increase of separation.
Figure \ref{2K} shows the detection advantage as a function of prior probability and weighting.
One can find that the performance advantage increases with the decrease of weighting.
These results originate from the fact that large separation or small weighting leads to a more significant difference between the two hypotheses, which is favourable to detection. 
The performance advantage is not significant ($<5\%$) when the weighting is greater than 0.9, which typically exists in most observations of extrasolar planets \cite{doi:10.1073/pnas.1304213111,annurev-astro-081309-130837}.
For a tiny separation such as $k\le 0.01$, the performance advantage is negligible ($<0.5\%$).

In particular, there exists a forbidden region satisfying
\begin{align}
P_1 < \frac{q}{1+q}.
\label{e15}
\end{align}
This is stemmed from the result that all eigenvalues are positive.
The detailed calculation in terms of sequential principal minor can be found in Appendix \ref{A}.
It should be noted that although the performance advantage without prior information ($P_1=0.5$) is greater than that with prior information ($P_1 \neq 0.5$), the opposite happens when we consider the quantum-optimal error probability.

Now we direct our attention on a symmetric scenario.
At this time, quantum states under two hypotheses can be written as 
\begin{align}
{\rho _1} = \left| {{\psi _0}} \right\rangle \left\langle {{\psi _0}} \right|,
\label{e16}
\end{align}
\begin{align}
{\rho _2} = q\left| {{\psi _ + }} \right\rangle \left\langle {{\psi _ + }} \right| + \left( {1 - q} \right)\left| {{\psi _ - }} \right\rangle \left\langle {{\psi _ - }} \right|.
\label{e17}
\end{align}
These two quantum states are composed by three non-orthogonal ket vectors.
As a consequence, we need at least a three-dimensional matrix to represent ${{P_2}{\rho _2} - {P_1}{\rho _1}}$.
In particular, these three ket vectors satisfy Eq. (\ref{e6}) with $\left\{ {i,j} \right\} \in \left\{ {0, \pm } \right\}$,
\begin{align}
\left\langle {x}
{\left | {\vphantom {x {{\psi _ \pm }}}}
	\right. \kern-\nulldelimiterspace}
{{{\psi _ \pm }}} \right\rangle  = {\left( {\frac{1}{{2\pi {\sigma ^2}}}} \right)^{{1 \mathord{\left/
				{\vphantom {1 4}} \right.
				\kern-\nulldelimiterspace} 4}}}\exp \left[ { - \frac{{{{\left( {x \pm {d \mathord{\left/
								{\vphantom {d 2}} \right.
								\kern-\nulldelimiterspace} 2}} \right)}^2}}}{{4{\sigma ^2}}}} \right],
\label{e18}
\end{align}
and $\left\langle {x}
{\left | {\vphantom {x {{\psi _0}}}}
	\right. \kern-\nulldelimiterspace}
{{{\psi _0}}} \right\rangle$ in Eq. (\ref{e7}).

Based on the Schmidt’s orthogonalization, we get three orthonormalized ket vectors as follows
\begin{align}
\left| 0 \right\rangle  =& \left| {{\psi _0}} \right\rangle, \\
\left| 1 \right\rangle  =& \frac{1}{{\sqrt {1 - {{\left| {{\delta _1}} \right|}^2}} }}\left( {\left| {{\psi _ + }} \right\rangle  - {\delta _1}\left| 0 \right\rangle } \right),\\
\left| 2 \right\rangle  =& \frac{1}{{\sqrt {1 - {{\left| {{\delta _2}} \right|}^2} - {{\left| {{\delta _3}} \right|}^2}} }}\left( {\left| {{\psi _ - }} \right\rangle  - {\delta _2}\left| 0 \right\rangle  - {\delta _3}\left| 1 \right\rangle } \right), 
\label{e21}
\end{align}
with
\begin{align}
{\delta _1} =& \left\langle {{{\psi _0}}}
{\left | {\vphantom {{{\psi _0}} {{\psi _ + }}}}
	\right. \kern-\nulldelimiterspace}
{{{\psi _ + }}} \right\rangle, \\
{\delta _2} =& \left\langle {{{\psi _0}}}
{\left | {\vphantom {{{\psi _0}} {{\psi _ - }}}}
	\right. \kern-\nulldelimiterspace}
{{{\psi _ - }}} \right\rangle, \\
{\delta _3} =& \frac{1}{{\sqrt {1 - {{\left| {{\delta _1}} \right|}^2}} }}\left( {\left\langle {{{\psi _ + }}}
	{\left | {\vphantom {{{\psi _ + }} {{\psi _ - }}}}
		\right. \kern-\nulldelimiterspace}
	{{{\psi _ - }}} \right\rangle  - \delta _1^ * \delta _2^{}} \right).
\label{e24}
\end{align}

Using Eq. (\ref{e9}), we can calculate and rewrite the above parameters as
\begin{align}
{\delta _1} =& {\delta _2} = \exp \left( { - \frac{{{k^2}}}{{32}}} \right) \equiv \delta, \\
\label{}
{\delta _3} =&  - {\delta ^2}\sqrt {1 - {\delta ^2}} 
\label{e26}
\end{align}
with $k = d/\sigma$.
Further, in the representation composed of $\left| 0 \right\rangle$, $\left| 1 \right\rangle$ and $\left| 2 \right\rangle$, two quantum states turn out to be
\begin{align}
{\rho _1} = \left[ {\begin{array}{*{20}{c}}
	1&0&0\\
	0&0&0\\
	0&0&0
	\end{array}} \right],
\label{e27}
\end{align}
\begin{widetext}
\begin{align}
{\rho _2} = q\left[ {\begin{array}{*{20}{c}}
	{{\delta ^2}}&{\delta \sqrt {1 - {\delta ^2}} }&0\\
	{\delta \sqrt {1 - {\delta ^2}} }&{1 - {\delta ^2}}&0\\
	0&0&0
	\end{array}} \right] + ( {1 - q} )\left[ {\begin{array}{*{20}{c}}
	{{\delta ^2}}&{ - {\delta ^3}\sqrt {1 - {\delta ^2}} }&{\delta \sqrt {1 - {\delta ^2} - {\delta ^4}\left( {1 - {\delta ^2}} \right)} }\\
	{ - {\delta ^3}\sqrt {1 - {\delta ^2}} }&{{\delta ^4}\left( {1 - {\delta ^2}} \right)}&{ - {\delta ^2}\left( {1 - {\delta ^2}} \right)\sqrt {1 - {\delta ^4}} }\\
	{\delta \sqrt {1 - {\delta ^2} - {\delta ^4}\left( {1 - {\delta ^2}} \right)} }&{ - {\delta ^2}\left( {1 - {\delta ^2}} \right)\sqrt {1 - {\delta ^4}} }&{1 - {\delta ^2} - {\delta ^4}\left( {1 - {\delta ^2}} \right)}
	\end{array}} \right].
\label{e28}
\end{align}
\end{widetext}
Based on these results, the quantum-optimal error probability can be calculated from Eq. (\ref{e2}).

\begin{figure*}[htbp]
	\centering
	\includegraphics[width=0.3\textwidth]{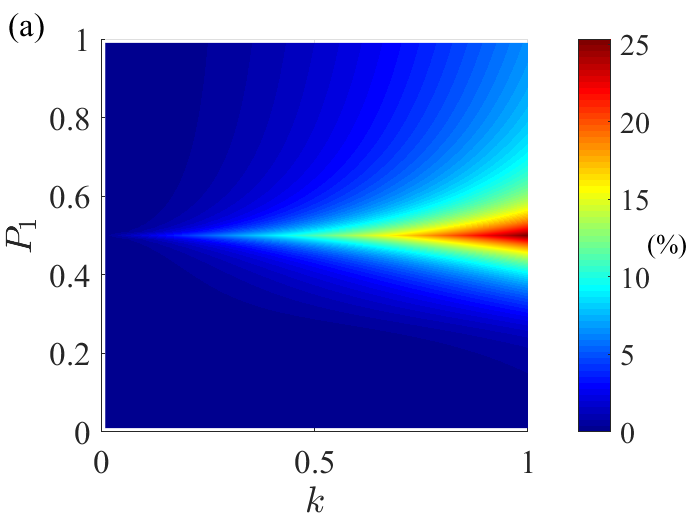}	\includegraphics[width=0.3\textwidth]{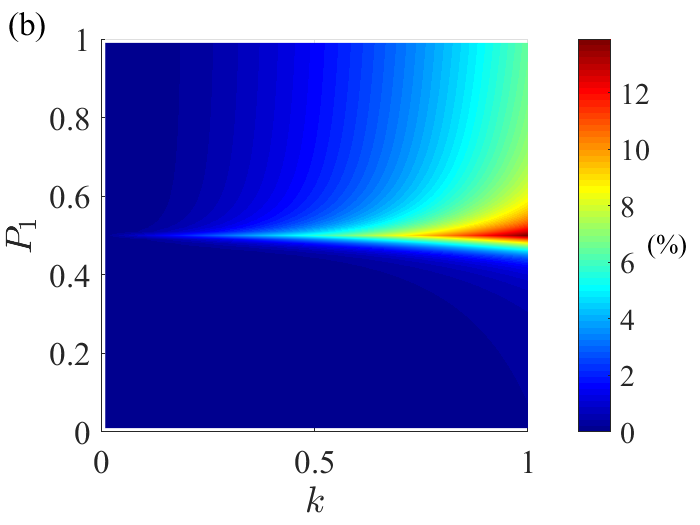}
	\includegraphics[width=0.3\textwidth]{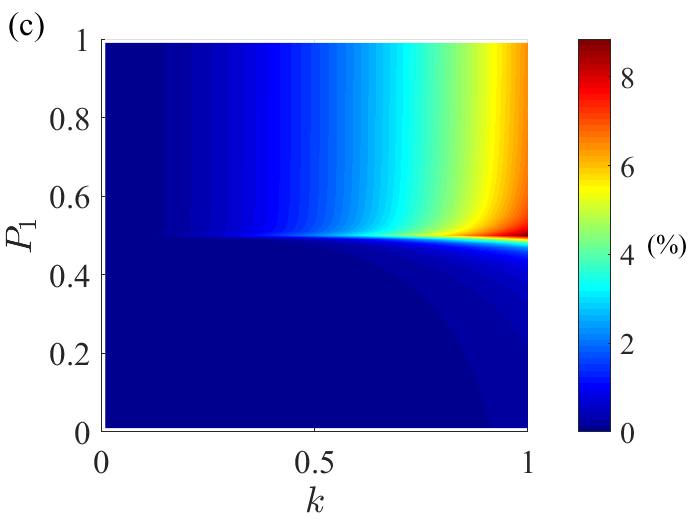}
	\caption{One-shot performance advantage as a function of prior probability and separation in a symmetric scenario, (a) $q = 0.1/0.9$; (b) $q = 0.3/0.7$; (c) $q = 0.5$. The color bars are displayed in percentage form.}
	\label{1Q}
\end{figure*}

\begin{figure*}[htbp]
	\centering
	\includegraphics[width=0.3\textwidth]{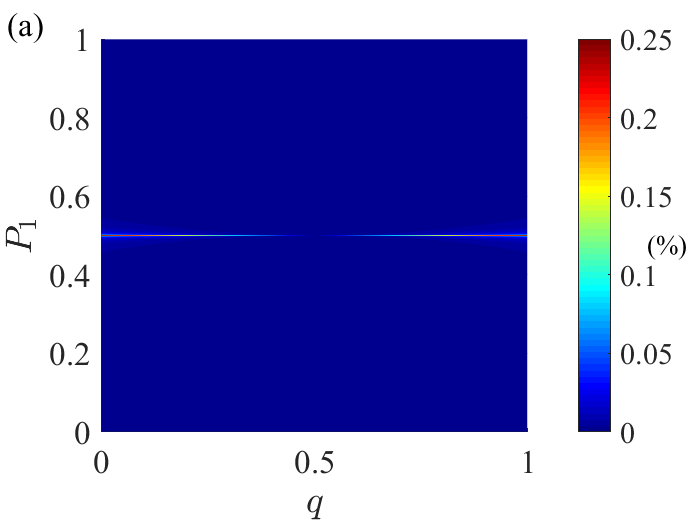}
	\includegraphics[width=0.3\textwidth]{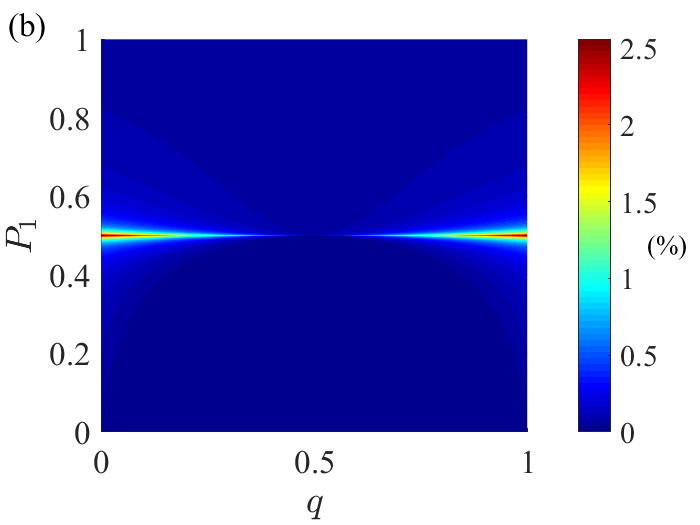}
	\includegraphics[width=0.3\textwidth]{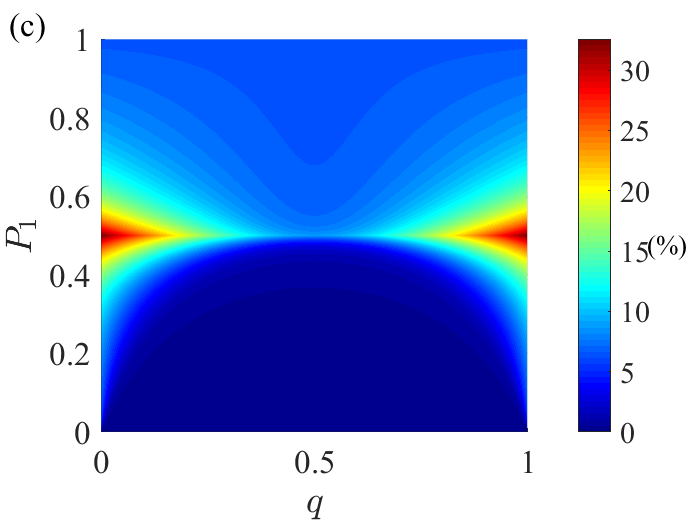}
	\caption{One-shot performance  advantage as a function of prior probability and weighting in a symmetric scenario, (a) $k = 0.01$; (b) $k = 0.1$; (c) $k = 1$. The color bars are displayed in percentage form.}
	\label{1K}
\end{figure*}

In Fig. \ref{1Q}, we give the performance advantage versus prior probability and separation. 
The result indicates that, for a fixed prior probability, the performance advantage is increased with increasing the separation.
Large spatial separation leads to an increase in the spatial size and gives movement of the center of two sources.
In Fig. \ref{1K}, the performance advantage against prior probability and separation is exhibited.
The performance advantage is symmetric with respect to $q=0.5$ and reaches its minimal value at $q=0.5$, regardless of separation.
This suggests that two equal-brightness sources are the most difficult to discriminate since the center of the two sources is still at 0.
Related to this, the performance advantage is not significant ($<3\%$) for $k<0.5$.
Regarding the unequal brightnesses, this center gradually moves towards $d/2$ or $-d/2$.
For a tiny separation such as $k\le 0.01$, the performance advantage can be neglected ($<0.25\%$). 
In particular, there is no forbidden region in this scenario, which is also proof in Appendix \ref{B}.

The previous analysis indicates that there are many similarities in the two scenarios with respect to performance advantage.
On the one hand, the performance advantage increases with the increase of separation.
On the other hand, the performance advantage with $P_1>0.5$ is generally superior to that with $P_1<0.5$.
These are due to the fact that large separation or value of $P_1$ increases the difference between the two hypotheses, which provides more detectable information.
In addition, there is a main difference between the performance advantages of the two scenarios.
In generally, the performance advantage of asymmetric scenario is greater than that of symmetric scenario.
This is mainly because the center of the two sources in the asymmetric scenario is near $d/2$ whereas that in the symmetric model is near $0$.
To be specific, there is a more significant difference between the two hypotheses in the asymmetric scenario.

\section{Quantum bound on multi-shot test}
\label{s4}

In general, making a decision after one-shot test is mainly used in proof-of-principle experiments \cite{PhysRevLett.127.040504}.
This originates from that the result of multiple one-shot tests can be used to reconstruct the probabilities of various outcomes and multiple decision results can be used to calculate the error probability.
In most of realistic scenarios, making a decision after multi-shot test is more useful.
In this section, we analyze $M$-shot test and discuss the relevant performance advantage.

\begin{figure*}[htbp]
	\centering
\includegraphics[width=0.3\textwidth]{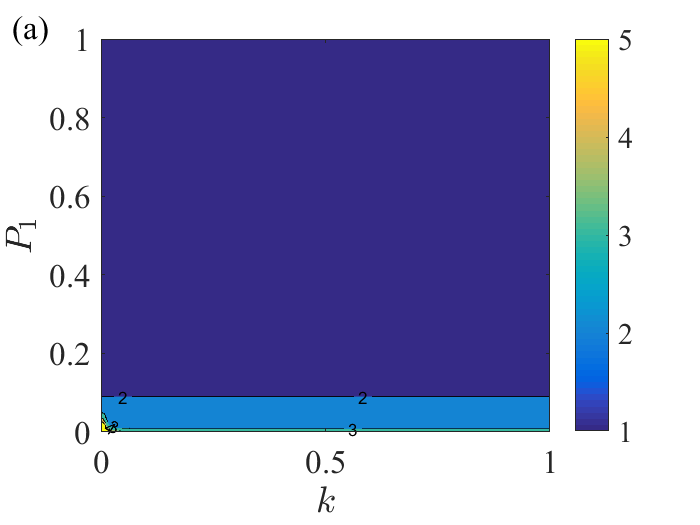}
\includegraphics[width=0.3\textwidth]{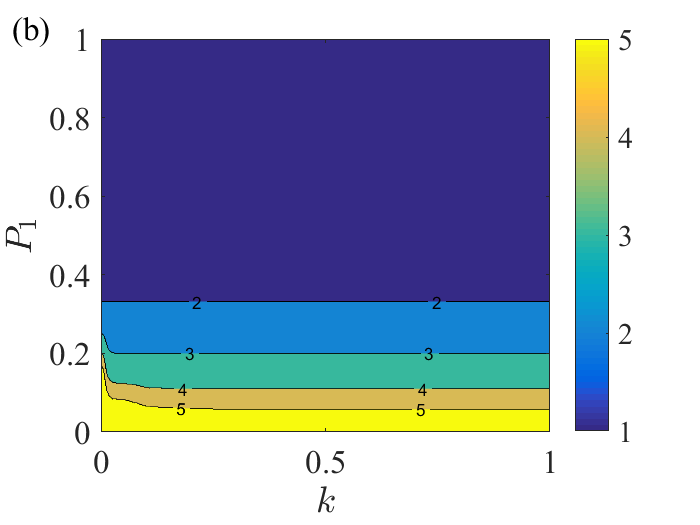}
\includegraphics[width=0.3\textwidth]{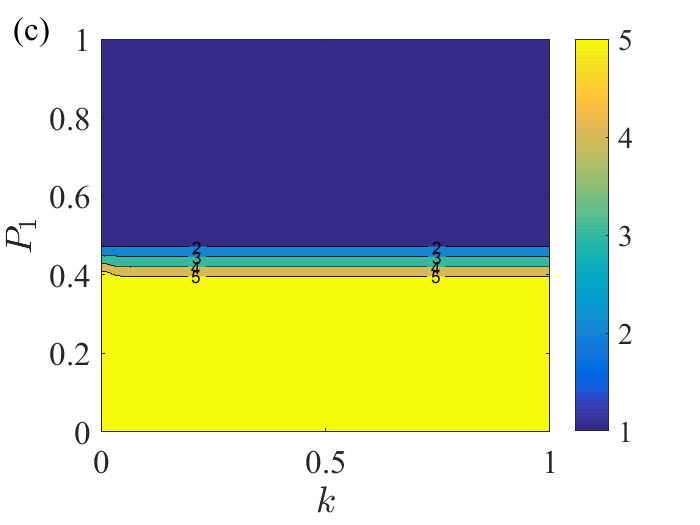}
\caption{The minimal value of $M$ achieving performance advantage as a function of prior probability and separation, (a) $q = 0.1$; (b) $q = 0.5$; (c) $q = 0.9$. Five regions correspond to $M=1$, $M=2$, $M=3$, $M=4$ and $M \ge 5$, respectively.}
\label{f6}
\end{figure*}

\begin{figure*}[htbp]
\centering
\includegraphics[width=0.3\textwidth]{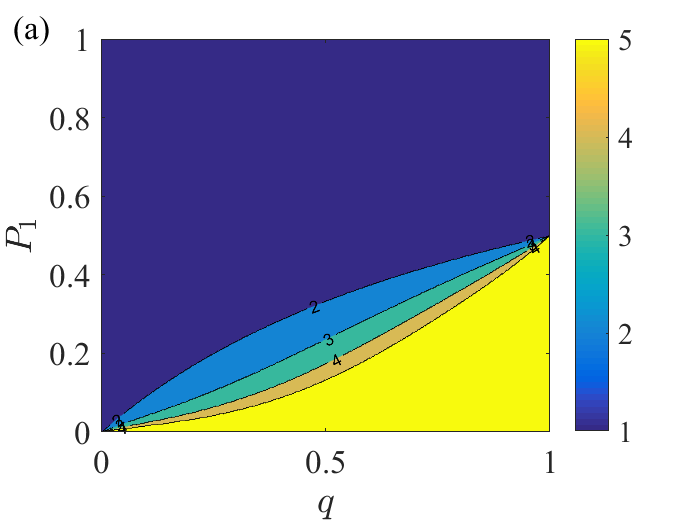}
\includegraphics[width=0.3\textwidth]{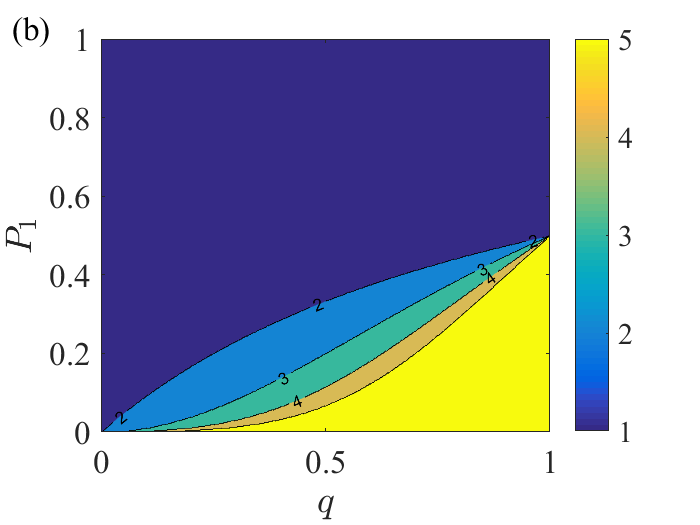}
\includegraphics[width=0.3\textwidth]{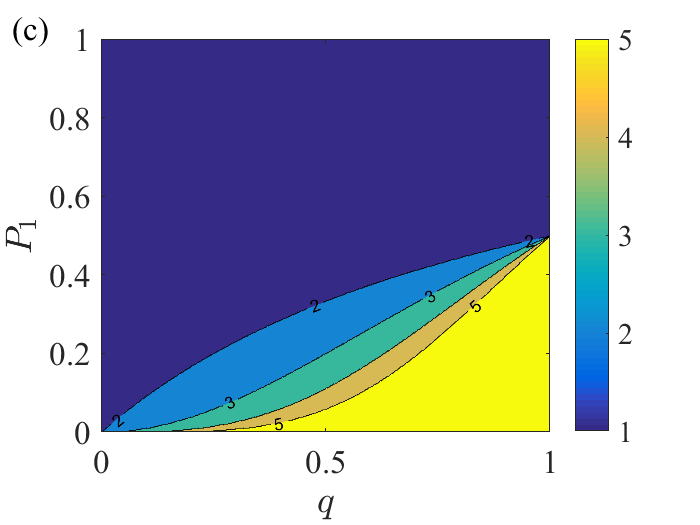}
\caption{The minimal value of $M$ achieving performance advantage as a function of prior probability and weighting, (a) $k = 0.01$; (b) $k = 0.1$; (c) $k = 1$. Five regions correspond to $M=1$, $M=2$, $M=3$, $M=4$ and $M \ge 5$, respectively.}
\label{f7}
\end{figure*}

For a small value of $M$, the corresponding quantum Helstrom bound can be directly calculated according to Eq. (\ref{e3}).
The previous section shows that the quantum-optimal error probability of making a decision after one-shot test is always superior to that of direct guess in a symmetric scenario; however, within certain parameter region in an asymmetric scenario, any decision after one-shot test cannot surpass direct guess.
For this reason, here we focus on the forbidden region in the asymmetric scenario and discuss the performance advantage of making a decision after $M$-shot test.
In Figs. \ref{f6} and \ref{f7}, we show the minimal value of $M$ required to decision surpassing direct guess. 
It turns out that one can break through the forbidden region by increasing the number of tests used in each decision.
Meanwhile, the value of $M$ decreases with increasing the brightness of the second source or the separation between the two sources.
This suggests that, as the difference between two hypotheses gets larger, the quantum-optimal error probability becomes lower.
As a matter of fact, the quantum-optimal error probability of making a decision after multi-shot test could always surpass the error probability of direct guess as long as two prior probabilities are nondegenerate.

For a large vaule of $M$, it is difficult to calculate the quantum Helstrom bound due to rapid expansion of matrix dimensionality.
Related to this, one generally settles for analyzing the asymptotic behavior with $M \gg 1$.
To be specific, the error probability in Eq. (\ref{e3}) decreases exponentially in $M$: 
\begin{align}
{E_{\min }} \sim \frac{1}{2}\exp \left( { - M{\xi _{\rm{Q}}}} \right),
\label{e29}
\end{align}
which is known as the quantum Chernoff bound with the relevant error exponent
\begin{align}
{\xi _{\rm{Q}}} =  - \ln \mathop {\min }\limits_{0 \le s \le 1} {\rm{Tr}}\left( {\rho _1^s\rho _2^{1 - s}} \right).
\label{e30}
\end{align}
For our model, it is not difficult to find that the $s$-overlap trace reaches its minimal value at $s = 0$. 
After straightforward calculation, the error exponents in asymmetric and symmetric scenarios are found to be
\begin{align}
{\xi _{{\rm{Qa}}}} =  - \ln \left[ {\frac{1}{2} + \frac{1}{2}\exp \left( { - \frac{{{k^2}}}{4}} \right)} \right]
\label{e31}
\end{align}
and
\begin{align}
{\xi _{{\rm{Qs}}}} = \frac{{{k^2}}}{{16}}.
\label{e32}
\end{align}

\section{Nearly optimal strategy along with decision rule}
\label{s5}
In the previous sections, we analyze the quantum-optimal error probability against prior probability and weighting.
In theory, a ready-made optimal detection is composed of  
the projectors onto the positive and negative supports of $\left\| {{\rho ^{\otimes M}_2} - {\rho^{\otimes M} _1}} \right\|/2$ \cite{yung2020one}; meanwhile, the optimal decision rule is likelihood ratio test \cite{van2004detection}.
However, the practical implementation of the optimal detection strategy is generally complicated, and the optimal decision rule is dependent on the true values of all parameters.
As a consequence, in a realistic scenario one needs to find a specific protocol including detection strategy and decision rule.

Here we consider the most general situation with $P_1 = P_2 =0.5$ and take $q = 0.5$ as an example.
As the first part of our protocol, the detection strategy is inspired by SLIVER and makes some improvements.
To be specific, photons are decomposed into odd and even spatial modes via SLIVER.
On the basis of this, we use two detectors to obtain these two mode probabilities.
Without loss of generality, we assume that the first (second) detector is used to record even (odd) spatial mode.

In an asymmetric scenario, even and odd spatial mode probabilities under $H_1$ hypothesis are given by
\begin{align}
{\rm{Pr}_{\rm a}}({\rm{On,Off}}) = &1, 
\label{e33} \\
{\rm{Pr}_{\rm a}}({\rm{Off,On}}) = & 0,
\label{e34}
\end{align}
while those under $H_2$ hypothesis are given by
\begin{align}
{\rm{Pr}_{\rm a}}({\rm{On,Off}}) = &\frac{1 }{4} \left [ 3+\exp \left( { - \frac{{{k^2}}}{{2}}} \right) \right ], 
\label{e35} \\
{\rm{Pr}_{\rm a}}({\rm{Off,On}}) = & \frac{1}{4} \left [ 1-\exp \left( { - \frac{{{k^2}}}{{2}}} \right) \right ],
\label{e36}
\end{align}
the details can be found in Appendix \ref{C}.
Similarly, in a symmetric scenario, even and odd spatial mode probabilities under $H_1$ hypothesis read
\begin{align}
{\rm{Pr}_{\rm s}}({\rm{On,Off}}) = &1, 
\label{e37}\\
{\rm{Pr}_{\rm s}}({\rm{Off,On}}) = & 0,
\label{e38}
\end{align}
while those under $H_2$ hypothesis are found to be 
\begin{align}
{\rm{Pr}_{\rm s}}({\rm{On,Off}}) = & \frac{1 }{2} \left [ 1+\exp \left( { - \frac{{{k^2}}}{{8}}} \right) \right ], 
\label{e39}   \\
{\rm{Pr}_{\rm s}}({\rm{Off,On}}) = & \frac{1}{2} \left [ 1-\exp \left( { - \frac{{{k^2}}}{{8}}}\right)\right ].
\label{e40}
\end{align}

Based on above probabilities, we can determine the decision rule, the second part of our protocol.
If the second detector is triggered by a single photon, then we accept $H_2$ hypothesis, otherwise $H_1$ hypothesis is accepted.
Then we find that the type-I error probability $\alpha$ (imagining the second source) and type-II error probability $\beta$ (missing the second source) satisfy
\begin{align}
\alpha = &0, \\
\beta = &{\rm{Pr}_{\rm {s/a}}}({\rm{On,Off}}),
\label{e42}
\end{align}
where the specific error probabilities are given by Eqs. (\ref{e35}) and (\ref{e39}).
Accordingly, the total error probability turns out to be 
\begin{align}
P_{\rm{err}} = \frac{1}{2}(\alpha + \beta) = \frac{1}{2}{\rm{Pr}_{\rm {s/a}}}({\rm{On,Off}}).
\label{e43}
\end{align}

\begin{figure}[htbp]
	\centering
	\includegraphics[width=0.3\textwidth]{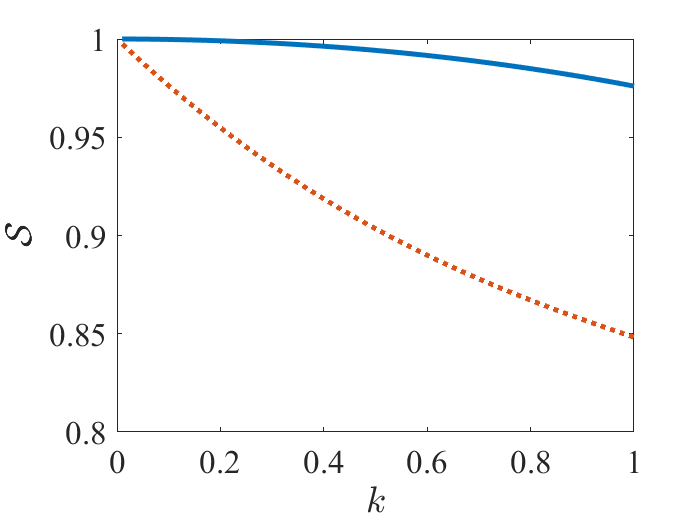}	
	\caption{The saturation in symmetric (solid line) and asymmetric (dotted line) scenarios as a function of separation with different total photon numbers.}
	\label{f8}
\end{figure}

To characterize the optimality of our protocol in one-shot test, we define 
\begin{align}
{\cal S} = \frac{E_{\min}}{P_{\rm{err}}}
\label{e44}
\end{align}
as saturation.
For comparison, we show the saturation as a function of separation in Fig. \ref{f8}.
It can be seen that, both in the symmetric and asymmetric scenarios, the saturation approaches 1 when the separation tends towards 0.
That is, our protocol is nearly optimal for a small separation.
Even for a separation that is not small enough, it can achieve saturation over 85\% within sub-Rayleigh region.
Significantly, our protocol performs better in the symmetric scenario when compared with the asymmetric scenario.

When we use $M$-shot test to make a decision, the error probability of our protocol is given by
\begin{align}
P_{\rm{err}} = \frac{1}{2}{{\rm{Pr}}^{M}_{\rm {s/a}}}({\rm{On,Off}}).
\label{e45}
\end{align}
This is due to the fact that we mistakenly accept $H_1$ hypothesis only when the second detector is not triggered during the entire detection process.
In other words, as long as there exists at least a detection event where the second detector is triggered, we can accept $H_1$ hypothesis, which is a correct decision.

For the convenience of comparison with the quantum Chernoff bound, the error probability in Eq. (\ref{e45}) can be rewritten as
\begin{align}
{P_{{\rm{err}}}} = \frac{1}{2}{\rm{exp}}\left( { - M{\xi _{{\rm{SLIVERs/a}}}}} \right)
\label{e46}
\end{align}
with the relevant error exponent
\begin{align}
{\xi _{{\rm{SLIVERs/a}}}} = - {\rm{lnPr}}_{{\rm{s/a}}}^{}({\rm{On}},{\rm{Off}}).
\label{e47}
\end{align}
The specific error exponents in the asymmetric and symmetric scenarios are given by
\begin{align}
{\xi _{{\rm{SLIVERa}}}} =&  - \ln \left[ {\frac{3}{4} + \frac{1}{4}\exp \left( { - \frac{{{k^2}}}{2}} \right)} \right],  \\
{\xi _{{\rm{SLIVERs}}}} =&  - \ln \left[ {\frac{1}{2} + \frac{1}{2}\exp \left( { - \frac{{{k^2}}}{8}} \right)} \right].
\label{e49}
\end{align}

\begin{figure}[htbp]
\centering
\includegraphics[width=0.3\textwidth]{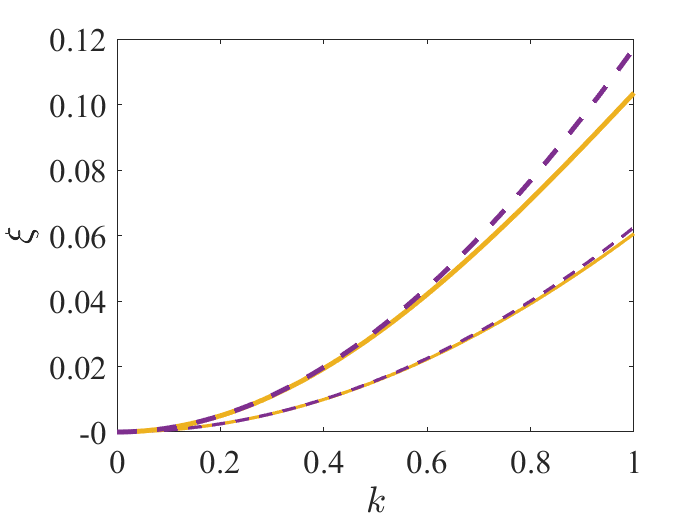}	
\caption{The error exponent as a function of separation. Meaning of various line styles: quantum Chernoff exponents in symmetric (thin dashed line) and asymmetric (thick dashed line) scenarios; error exponents of SLIVER in symmetric (thin solid line) and asymmetric (thick solid line) scenarios.}
\label{f9}
\end{figure}

Figure \ref{f9} shows error exponent of the quantum Chernoff bound and that of modified SLIVER in asymmetric and symmetric scenarios. 
One can find that two error exponents are almost equal when the separation satisfies $k < 0.4$ ($k < 0.7$) in the asymmetric (symmetric) scenario.
This result means that our protocol can work well within sub-Rayleigh region, specially for very small separation, since the relevant error probability can approach the quantum Chernoff bound.
As a matter of fact, our protocol is nearly optimal for one-shot test as well as multi-shot test.

\begin{figure}[htbp]
\centering
\includegraphics[width=0.3\textwidth]{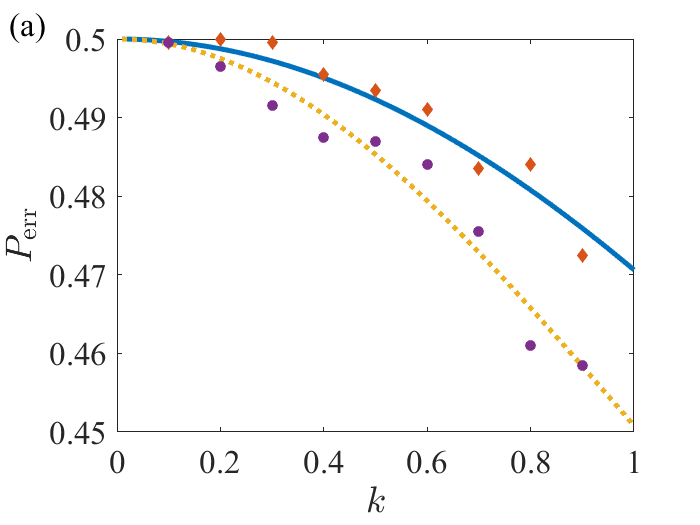}	
\includegraphics[width=0.3\textwidth]{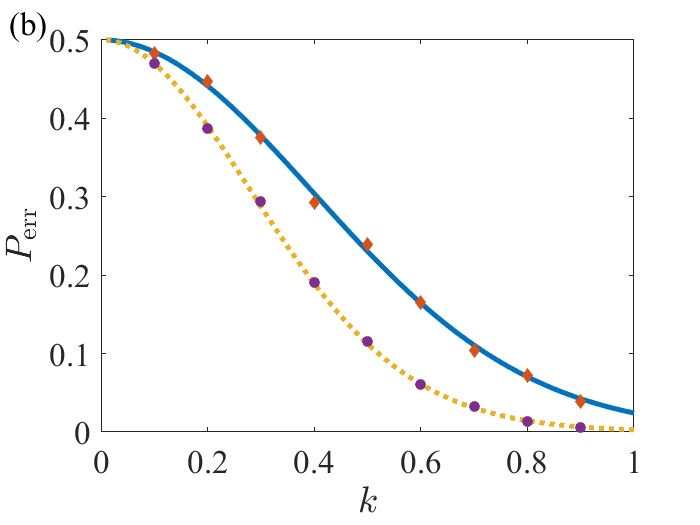}	
\caption{The error probability as a function of separation with (a) one-shot test ($M = 1$) and (b) multi-shot test ($M = 50$). Meaning of various line styles and symbols: theoretical result (solid line) and simulation result (diamonds) in an asymmetric scenario; theoretical result (dotted line) and simulation result (dots) in a symmetric scenario.}
\label{f10}
\end{figure}

Figure \ref{f10} provides numerical simulation based on Monte Carlo method.
For each of two hypotheses, we repeat the one-shot or multi-shot test 1000 times.
The error probability is equal to the ratio of total number of wrong decisions to the total iteration number.
It can be seen that simulation and theory are in close agreement in both scenarios.
In contrast, multi-shot test exhibits excellent capability to reduce the error probability, and the agreement between theory and simulation in multi-shot test is better than that in one-shot test.
With the increase of separation, the error probability of multi-shot test diminishes at a very fast rate.

It is beneficial to compare our protocol with other symmetric hypothesis testing protocols related to this subject.
In particular, Ref. \cite{lu2018quantum} presented a SLIVER-based protocol in a symmetric scenario, which performs slightly better than our protocol and is saturated by the quantum Chernoff bound throughout sub-Rayleigh region.
The expense of this advantage is assumption of a known total source brightness, which is a somewhat restrictive constraint on the applicability.
On the other hand, a protocol similar to SLIVER was reported in Ref. \cite{Wadood24} and also showed a slight advantage over our protocol.
It is not surprising for such advantage since this protocol presumes that the true values of all parameters are known, a situation that seldom occurs in realistic scenarios.

\section{Discussion and outlook}
\label{s6}

Although our protocol is theoretically analyzed and demonstrates a nearly quantum-optimal error probability, there are two important explanations to our work. 
Firstly, our protocol can be implemented under the current experimental techniques. 
Let us consider the task of discriminating whether there is a potential planet around a star of interest, which is an instance that can be analogized to an asymmetric case. 
Given that the position of a star is generally known in advance, we can direct the telescope at the star.
At this point, the star remains at the origin of the image plane regardless of the presence of the planet. 
Then we can deploy a coaxial Mach-Zehnder interferometer behind the telescope.
On this basis, an optical image-inversion device, e.g. a Dove prism, is embedded with one arm of the interferometer.
In doing so, photons in odd and even spatial modes are detected from the two output ports of the interferometer, respectively. 
Two Geiger mode avalanche photodiodes can be used to detect photons.

Secondly, there are still some potential challenges in realistic application scenarios, which can be roughly divided into two types. 
The first type of challenge mainly comes from the imperfection of optical devices, like detector efficiency, dark counts, and cross talk. 
The second type of challenge mainly stems from non-cooperative detection objects including fluctuation in source brightness and variation in separation. 
In general, one can effectively circumvent the first type of challenge with pre-calibration. 
For example, detector efficiency, rate of dark counts, and rate of cross talk can be determined through a proof-of-principle experiment. This method has been confirmed in some experiments, which greatly reduces the undesirable effect \cite{Tang:16,Larson2019,Santamaria:24,Wadood24}. 
With respect to the second type of challenge, an effective method is to use the result of each test to update the posterior probability and then to dynamically adjust the detection strategy. 
Despite some explorations on this subject \cite{Bonsma-Fisher_2019,Grace:20,PhysRevA.104.022410}, a real-time and efficient adaptive algorithm is still an open question.
This may be the most promising direction as it provides a powerful and flexible tool for smart quantum imaging.

\section{Conclusion}
\label{s7}
In summary, we addressed the task of discriminating one-versus-two incoherent sources based on quantum hypothesis testing.
The quantum lower bound on error probability in each decision was reported in asymmetric and symmetric scenarios.
For one-shot test, we calculated the quantum Helstrom bound and compared it with the error probability of prior-based direct guess.
In a symmetric scenario, one-shot test is enough to make a decision that is superior to direct guess; however, such result no longer holds within certain parameter region in an asymmetric scenario.
With respect to this region, we determined the minimal number of tests required to a decision outperforming direct guess.
For multi-shot test, we showed the performance advantage of few-shot test and studied the asymptotic behavior for a large number of tests.
The quantum Chernoff bound was calculated, and analytical result was provided.
For practical purposes, we proposed a nearly optimal protocol based on modified SLIVER.
We can use this protocol to make a decision independent of various parameters, such as separation and brightness.
In each of two scenarios, the error probability can approach the quantum lower bound either in one-shot test or multi-shot test. 
As a complement, we discussed potential challenges and solutions in a realistic scenario.
Our results may possess contribution to practical applications such as fluorescence microscopy and astronomy imaging.

\section*{Acknowledgment} 
This work was supported by the Program of Zhongwu Young Innovative Talents of Jiangsu University of Technology (20230013).

\appendix
\section*{Appendix}

\section{Calculation for forbidden region with one-shot test in an asymmetric scenario}
\label{A}

In the asymmetric scenario, ${{P_2}{\rho _2} - {P_1}{\rho _1}}$ is a two-dimensional matrix.
In terms of sequential principal minor and notation $\Omega \equiv {{P_2}{\rho _2} - {P_1}{\rho _1}}$, all eigenvalues are positive if
\begin{align}
{\Omega _{11}} > 0,{\kern 3pt}  {\Omega _{11}}{\Omega _{22}} - {\Omega _{12}}{\Omega _{21}} > 0,
\label{A1}
\end{align}	
and all eigenvalues are negative if
\begin{align}
{\Omega _{11}} < 0,{\kern 3pt}  {\Omega _{11}}{\Omega _{22}} - {\Omega _{12}}{\Omega _{21}} > 0.
\label{A2}
\end{align}	
After some algebraic calculations, one can find that Eq. (\ref{A2}) is unsolvable and the solutions for Eq.
(\ref{A1}) is 
\begin{align}
P_1 < \frac{q}{1+q},{\kern 3pt}  P_1 < \frac{q+(1-q)\tau^2}{1+q+(1-q)\tau^2}.
\label{A3}
\end{align}
Finally, by combining the above equations, the solution for forbidden region is given by
\begin{align}
P_1 < \frac{q}{1+q}.
\label{A4}
\end{align}

\section{Calculation for forbidden region with one-shot test in a symmetric scenario}
\label{B}

In the symmetric scenario, $\Omega$ is a three-dimensional matrix.
In terms of sequential principal minor, all eigenvalues are positive if
\begin{align}
{\Omega _{11}} > 0,{\kern 3pt}  {\Omega _{11}}{\Omega _{22}} - {\Omega _{12}}{\Omega _{21}} > 0,{\kern 1pt} \det \left| \Omega  \right| > 0
\label{B1}
\end{align}	
and all eigenvalues are negative if
\begin{align}
{\Omega _{11}} < 0,{\kern 3pt}  {\Omega _{11}}{\Omega _{22}} - {\Omega _{12}}{\Omega _{21}} > 0,{\kern 1pt} \det \left| \Omega  \right| < 0
\label{B2}
\end{align}	

\begin{figure}[htbp]
	\centering
	\includegraphics[width=0.3\textwidth]{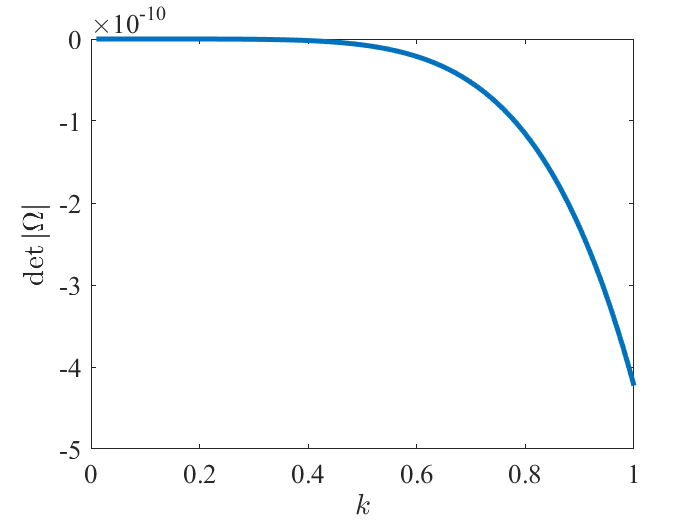}
	\caption{The determinant of $\Omega$ as a function of separation. Each point with a fixed $k$ is the maximal value over all values of $q$ and $P_1$. }
	\label{Proof}
\end{figure}

The solutions of the above inequalities involve solving a cubic equation.
Here we give an intuitive result since the analytical solution is cumbersome. 
As shown in Fig. \ref{Proof}, the determinant of $\Omega$ is always negative.
That is, we only need to focus on Eq. (\ref{B2}).
Further, we find that the first two inequalities in Eq. (\ref{B2}) are unsolvable.
Hence, there is no forbidden region in this scenario.

\section{Calculation for spatial mode probabilities in SLIVER}
\label{C}

Here we show the calculation for even and odd spatial mode probabilities in SLIVER.
For the convenience of analysis, we use the semi-classical method given by Ref. \cite{Nair:16}.
We assume that the point-spread function is normalized and circularly symmetric, i.e.,
\begin{align}
\int_{ - \infty }^\infty  {{{\left| {\psi \left( x \right)} \right|}^2}dx}  = 1
\label{C1}
\end{align}
and
\begin{align}
{\psi \left( x \right)}  = {\psi \left( -x \right)}.
\label{C2}
\end{align}	
Under $H_1$ hypothesis, there exists a single source, a completely even spatial mode in both scenarios.
Hence, we only need to analyze two scenarios under $H_2$ hypothesis.

For an asymmetric scenario, the complex amplitude of optical field on the image plane can be written as
\begin{align}
E\left( x \right) = {C_1}\psi \left( x \right) + {C_2}\psi \left( {x - d} \right).
\label{C3}
\end{align}	
The source amplitudes $C_1$ and $C_2$ are circular-complex Gaussian random variables, of which their first and second moments satisfy
\begin{align}
\overline {{C_i}}  = \overline {{C_i}{C_j}}  = \overline {C_i^ * {C_j}}  = 0
\label{C4}
\end{align}	
and
\begin{align}
\overline {C_i^ * {C_i}}  = \frac{\epsilon }{2}
\label{C5}
\end{align}	
with $i,j \in \left\{ {1,2} \right\}$.
As mentioned in the main text, two detectors record the even and odd spatial modes.
The complex amplitudes of these two modes are given by
\begin{align}
\nonumber {E_1} = & \frac{{E\left( x \right) + E\left( { - x} \right)}}{2}\\
= & \frac{1}{2}\left[ {2{C_1}\psi \left( x \right) + {C_2}\psi \left( {x - d} \right) + {C_2}\psi \left( {x + d} \right)} \right],\\
\nonumber {E_2} = &\frac{{E\left( x \right) - E\left( { - x} \right)}}{2}\\
=& \frac{1}{2}\left[ {{C_2}\psi \left( {x - d} \right) - {C_2}\psi \left( {x + d} \right)} \right].
\label{C7}
\end{align}	
Further, the mean photon numbers received by two detectors are given by 
\begin{align}
{{\bar N}_1} = &\int_{ - \infty }^\infty  {{{\left| {{E_1}} \right|}^2}dx}  = \frac{\epsilon }{4}\left[ {3 + \exp \left( { - \frac{{{k^2}}}{2}} \right)} \right],\\
{{\bar N}_2} = & \int_{ - \infty }^\infty  {{{\left| {{E_2}} \right|}^2}dx}  = \frac{\epsilon }{4}\left[ {1 - \exp \left( { - \frac{{{k^2}}}{2}} \right)} \right].
\label{C9}
\end{align}	
Finally, we can obtain the normalized spatial mode probabilities, Eqs. (\ref{e35}) and (\ref{e36}), in the main text.

For a symmetric scenario, by replacing the complex amplitude in Eq. (\ref{C3}) with
\begin{align}
E\left( x \right) = {C_1}\psi \left( {x - \frac{d}{2}} \right) + {C_2}\psi \left( {x + \frac{d}{2}} \right).
\label{C10}
\end{align}	
Through the same method, we can obtain the normalized spatial mode probabilities, Eqs. (\ref{e39}) and (\ref{e40}), in the main text.


%

\end{document}